# Proposing a New Method for Query Processing Adaption in DataBase

Mohammad-Reza Feizi-Derakhshi, Hasan Asil, Amir Asil

**Abstract**— This paper proposes a multi agent system by compiling two technologies, query processing optimization and agents which contains features of personalized queries and adaption with changing of requirements. This system uses a new algorithm based on modeling of users' long-term requirements and also GA to gather users' query datas. Experimented Result shows more adaption capability for presented algorithm in comparison with classic algorithms.

*Index Terms*— Queries Processing, Adaption, Agent.

———————————— ◆ ————————————

## 1 INTRODUCTION

In recent years, use of adaptive query processing (AQP) has been increased as a solution to the problems of query optimization and execution –across relational, text or XML data regardless of whether the data is accessed locally ,from the web or etc. Much of this is motivated by the emergency requirement of domains to optimization of query processing.

When Selinger-style query optimization failed, the result (systems flurry and creating new algorithms) made vendors like Microsoft, IBM and Oracle to investigate and deploy adaptivity features of their database products.

In this paper we attempt to articulate a "global view" covering existing techniques. Two particular dimensions we consider are planning space and the way of execution and optimization. We also discuss about benefits and drawbacks of various techniques and identify open research problems. Also we're going to present a way to increase adaptivity features of database and optimize it by agents and genetic algorithms.

Goal of database adaption is to select suitable data and ways of answering to queries according to users' long-term requirements of a dynamic and large collection of informative non-structural and semi-structural units.

## 2 AIM OF AQP

Declarative queries are a central value proposition of the relational model, letting the users specify only what results they want without having to worry about the strategy (plan) used to access and combine the data. Finding the best plan (query optimization) was addressed in even the first RDBMS- most successfully by Salinger's dynamic programming algorithm in system R. System R divided to optimization and execution. Over time this optimization approach has been improved (exploring more exhaustive plans, using histograms, adding cross-block query rewrites), but the basic system R architecture lives on in most query processors. Unfortunately, using such a method has extended in new and various fields, too, e.g. data streams, wide area data sources and interactive query environments that approach has run into limitations. But it's possible to extend query processing optimization in other ways and we are going to present a method which can be optimized by long- term adaptive query processing in database and genetic algorithm.

## 3 PREVIOUS NON SELINGER WORKS

Several techniques have been proposed to extend the query Optimization process to solve some of these problems:

1. Incorporating feedback from previous query executions for better selectivity/cardinality
2. Parametric techniques to systematically postpone making certain decisions as late as possible
3. Least excepted cost and by optimization techniques which dispenses the possibility

These techniques are based on static data details keeping and own limited amplitude but there are other techniques which propose new ways via adaptive techniques. The followings are two techniques in this field:

1. Selection ordering technique: selection ordering undertakes the way to exchange given set of commutative filters (selections) to all the tuples of a relation. In this technique new techniques are presented by using Greedy techniques and monitoring tuples properties continuously and adapting processes.
2. Adaptive join processing: the design and analysis of adaptive techniques for join queries is more complicated than selection ordering. The space of execution is much larger and more complex. This technique divides queries to :
- Independent pipelined executions

————————————————

- *Feizi is with the Department of Computer Science, University of Tabriz, Tabriz, Iran.*
- *Hasan Asil is with the Department of Computer, Islamic Azad University, Shabestar branch, Tabriz, Iran.*
- *Amir Asil is with Industrial Management Institute Azarbayjan, Tabriz, Iran.*

- Dependent pipelined executions
- Non- pipelined executions

And presents a solution for each one, But each one of pointed techniques has its own problem like parametric query optimization, parallelism and need for large memory of execution.

Query optimization allows us to reach the goals of optimization. Usually, methods of query processing optimization in database are divided into 3 groups:

1. Server: In this group of methods, it was tried to reach optimization goals by the changes performed on server, like hardware changes.

2. Query: in this group, optimization occurs by some techniques and changes performed on the query. One of these methods is changing parameters of a query.

3. Session: methods of this group do not apply in any of the above groups. Methods like paralleling, using cach place in this group.

In this research trying to present a new method base from the methods pointed out on the third group for optimizing query processing in database.

GA technique is a global repeatable optimization technique which receives guidance from natural ripening and genetics. GA has proved its successfulness in real world problem optimization which usually owns a very large working space. For this reason we can use genetic algorithm for reaching an optimized query.

## 4 SUGGESTED ALGORITHM

In this algorithm we suggest a multi agent system by compiling query optimizing and agents. This algorithm attempts to prepare a personalized environment for users and also attempts to propose style of query based on users' query by data gathering technology. Therefore we are engaged with query processing and the ways of gathering and answering for queries. For gathering data using genetic algorithms for producing a collection of optimized queries and by sending this queries and proposing method, answers for query by the previous data.

Database of software sends query according to user's need and replies based on these needs and queries. These softwares usually send queries to database, receive response and reply to the user's need. In these softwares sent queries were owned by the same structure and they are repeated as time goes by. Therefore we are going to present a method that database identifies queries of the same kind over a period of time. This method also identifies more repetitive queries sent to database and replies to this query by distinct execution plan. In other words we want to adapt database to process the queries by prepared execution plans faster and cheaper.

Suggested algorithm is based on vectorial space model. In this model query has seen as vector of an user in a Euclidean space and each user is one of its dimensions.

1. For discovering user weigh of each vector, weighing method is used. Weigh of ith user of jth vector discovers by following frame:

$$W_{ij} = (freq_{i,j} / \max freq_{i,j})*(\log N/n_i) \qquad (1)$$

- ✓ N: Types of queries in system
- ✓ $n_1$: number of queries that $K_1$th user appears in them
- ✓ $freq_{i,j}$: repeat times of $K_1$th user in Djth query
- ✓ $\max freq_{i,j}$: maximum times of repeat for all users of one query

2. In this model for comparing two vectors in the field of similarity, Cosine of pented angel between these vectors is needed.

$$Sim (D_j, Q_i) = (\sum_{k=1}^{T} d_{jk}.q_{ik})/ (\sum_{k=1}^{T} (d_{jk})^2) (\sum_{k=1}^{T} (q_{ik})^2) \qquad (2)$$

3. Cardinality is gent conversion of input queries to optimized query by using rate of users' queries from a particular type of query. Recho frame is one of common ways to calculate cardinality:

$$(3)$$

$$Q_{new}=\gamma Q_{old}+\beta/ |D_r|*(\lambda D_r\_/ |D_n|)*\sum D_j\_D_n D_j$$

At the pointed frame Qnew has created gently until it converges to optimized vector. Dr and Dn are two separate collection (Dr_Dn=Ø). And $\gamma$, $\beta$, $\lambda$ parameters show amount of partnership and effect of frame parameters (Qold, $D_1$ and $D_2$ collections) informing new vector (Qnew).

But this algorithm profile structure is in this shape that: "for each user one profile defines" for modeling of users' various queries, each profile has divided to some parties. Each party shows certain user queries and it showed as a T dimensions vector. (T: Number of users)

## 5 OPTIMIZATION PROCESS

Various queries are sent to database and there are some costs to pay on replying in each query. Usually sent orders are divided into four groups:

1. Select
2. Update
3. Delete
4. INSERT

Way of compare queries in database follows a distinct format and standard to constrain less cost when the queries are examined.

Each of these queries require cost to process and various methods have tried to reduce this expenditure.

Optimization process has expressed in suggested system of Algorithm 1. As it seems this process consists of comparing gathered query vectors and respective party and scoring to them.

Style of optimization of each query and its execution and also rate of users' request is effective in scoring and privileged technique for user query processing selects.

Input: collected Query
Output: Suggest Process List

1. Compare Each Process vector Dk whith its repected category
$c_i$, and then calculate its score
Score ($C_i$, $D_k$) =W ^C pass Sim (Desc C1…..)
2. Rank scored query in decrwasing order _
3. Select highest style to prcess query

**Algorithm 1 – Query processing optimization process**

Base of user's informational requirements learning process is applying user's cardinality about queries asked by user and also style of system answer for this queries. In fact, user's cardinality applies by updating its profile.

**Input: new query processor**
**Output: Updated Profile**

1. If new query ($D_{new}$) is provided by the user as sample query then
2. Begin
3. In Profile Pj for each category C1, compare $D_{new}$ with $c_i$.
Relevance ($c_1$, $D_{new}$) =Max {sim (Desc ^$C_1$ pass, $D_{new}$)}
4. $D_{new}$ Relates to the best matched Category $C_{best}$
$C_{best}$=Arg max {Relevance ($c_1$, $D_{new}$)}
5. End
6. Else
7. $C_{best}$=Query category
8. End if
9. If Relevance ($c_{Best}$, $D_{new}$) $\geq \theta$        then
10 .Update Existing Category (p1, Feedback Type,$\alpha$ , $D_{new}$)
11. Else
12. CreateNewCategory ($p_j$, feedback Type, $\alpha$ , $D_{new}$)
13. Endif

**Algorithm 2 – User's informational requirements learning Algorithm in suggested system**

In Algorithm 2 if the new query ($D_{new}$) has been sent to system by user as a sample, it compares the query with all of user
profiles, if similarity rate between Cbest and Dnew is less than a certain amount likes  then a new profile creates but if.

# 6 USING GENETIC ALGORITM IN QUERY CREATION

This technique uses a genetic algorithm to reach optimized queries.

In this problem users of one profile are supposed as searching space and the goal is to find an optimized syntax for users' queries. In this problem instead of coding a solu-tion as a bit string, genetic algorithm applies on solution and as we know solution consists of a vector of users. Also evaluation function defines as it comes below:

$$\text{Fitness (qk c1)} = (\textstyle\sum m \text{ j=1 } \phi \text{ (ci, Dj))/ m} \qquad (4)$$

$$\phi \text{ (ci, Dj) =sim (Descci pos, Dj)-Sim (DescCi neg,DJ)} \qquad (5)$$

At the pointed frame, eligibility and style of answering for query is average of $\phi$ (ci, dj) function for the number of m queries. This function is similar to style and way of answering for a query, with this difference that here weigh is not describer. Query is a non-Boolean expression and some users using it. Style of answering for queries is such that the system checks the query and scrutinizes group and user that asks the query and then selects a solution according to previous adaptations and the solution answered for that query and after that answers to user's query.

# 7 SYSTEM EVALUATION

This system has designed and simulated completely object oriented and the agents which have a rule in this system are: way of optimization and style of answering for query, creating profiles and updating them and execution of gathering algorithm and selecting way of answering them.

System evaluation process by executing some experiments which virtual user and queries are used in them, owns two main parameters. These parameters are: user model and style of answering for queries.

Requirement of this experiment is some queries and we use 8400 queries related to various relationships. Each query belongs with a group which has divided before. We suppose that the number of users in system is 160 per-sons.

For evaluating our accuracy and proficiency we suppose accuracy of learning ability of users' informational requirements evaluation and adapting with changes as our first goal. Way of algorithm applying simulates in real system. At the start profiles empty, then the profiles are categorized based on query and style of answering.

In next experiment rate of adaption and proficiency are supposed and experimental samples are shown as it comes below. As it clears, by increasing system working and increasing of queries adaption, proficiency of system increases in comparison to normal state.

Of course as we know it's possible some queries are asked system out of users' profiles but we know speed of genetic algorithm adaption in cases which sudden changing occurs is high and pointed system has ability of query adapting.

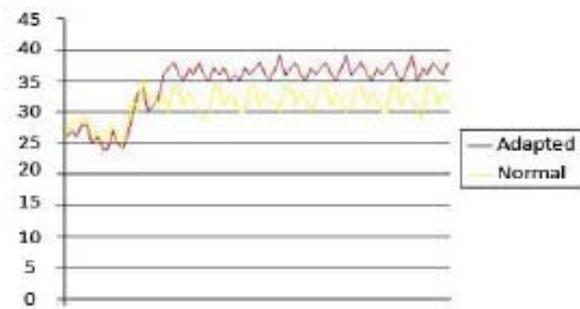

Fig.1. Proficiency of adapted algorithm

# 8 Result

In this paper it's tried to introduce a new algorithm for optimizing database query processing based on users' long-term queries. Using this technique has some benefits. The first one is existence of a clear model of requirements and users' queries, also style of their using of data banks and style of services answer to users' queries; also it prevents some repetitive operations in query processing.

Results of experiment show that suggested algorithm has good ability in modeling of users' queries and adapting database based on users requirement; also using genetic algorithms has improved and compilation of this technique with other calculating and learning methods like Nerve Plexus will considered.